\documentclass[9pt,twocolumn,twoside]{opticajnl}
\journal{opticajournal} 

\setboolean{shortarticle}{true}

\usepackage{graphicx,amsmath}

\usepackage{color}
\usepackage{siunitx}

\DeclareMathOperator{\Rop}{\mathcal{R}}

\DeclareMathOperator{\Pop}{\mathcal{D}}

\DeclareMathOperator{\Mop}{\mathcal{M}}

\title{Single-distance nano-holotomography with coded apertures}

\author[1,*]{Viktor Nikitin}
\author[2]{Marcus Carlsson}
\author[3]{Rajmund Mokso}
\author[4]{Peter Cloetens}
\author[1]{Do\u{g}a G\"ursoy}

\affil[1]{Advanced Photon Source, Argonne National Laboratory, 9700 S Cass Ave, Lemont, IL 60439, USA}
\affil[2]{Centre for Mathematical Sciences, Lund University, Sölvegatan 18, 223 62 Lund, Sweden}
\affil[3]{Department of Physics, Danish Technical University, Fysikvej, 310, 2800 Kgs. Lyngby, Denmark}
\affil[4]{ESRF-The European Synchrotron 71, Avenue des Martyrs, 38043 Grenoble, France}

\affil[*]{vnikitin@anl.gov}

\begin{abstract}
High-resolution phase-contrast 3D imaging using nano-holotomography typically requires collecting multiple tomograms at varying sample-to-detector distances, usually 3 to 4. This multi-distance approach significantly limits temporal resolution, making it impractical for \textit{operando} studies. Moreover, shifting the sample complicates reconstruction, requiring precise alignment and interpolation to correct for shift-dependent magnification on the detector. In response, we propose and validate through simulations a novel single-distance approach that leverages coded apertures to structure beam illumination while the sample rotates. This approach is made possible by our proposed joint reconstruction scheme, which integrates coded phase retrieval with 3D tomography. This scheme ensures data consistency and achieves artifact-free reconstructions from a single distance.
\end{abstract}

\setboolean{displaycopyright}{false} 

\begin{document}

\maketitle

\begin{figure}[ht!]
\centering\includegraphics[width=0.375\textwidth]{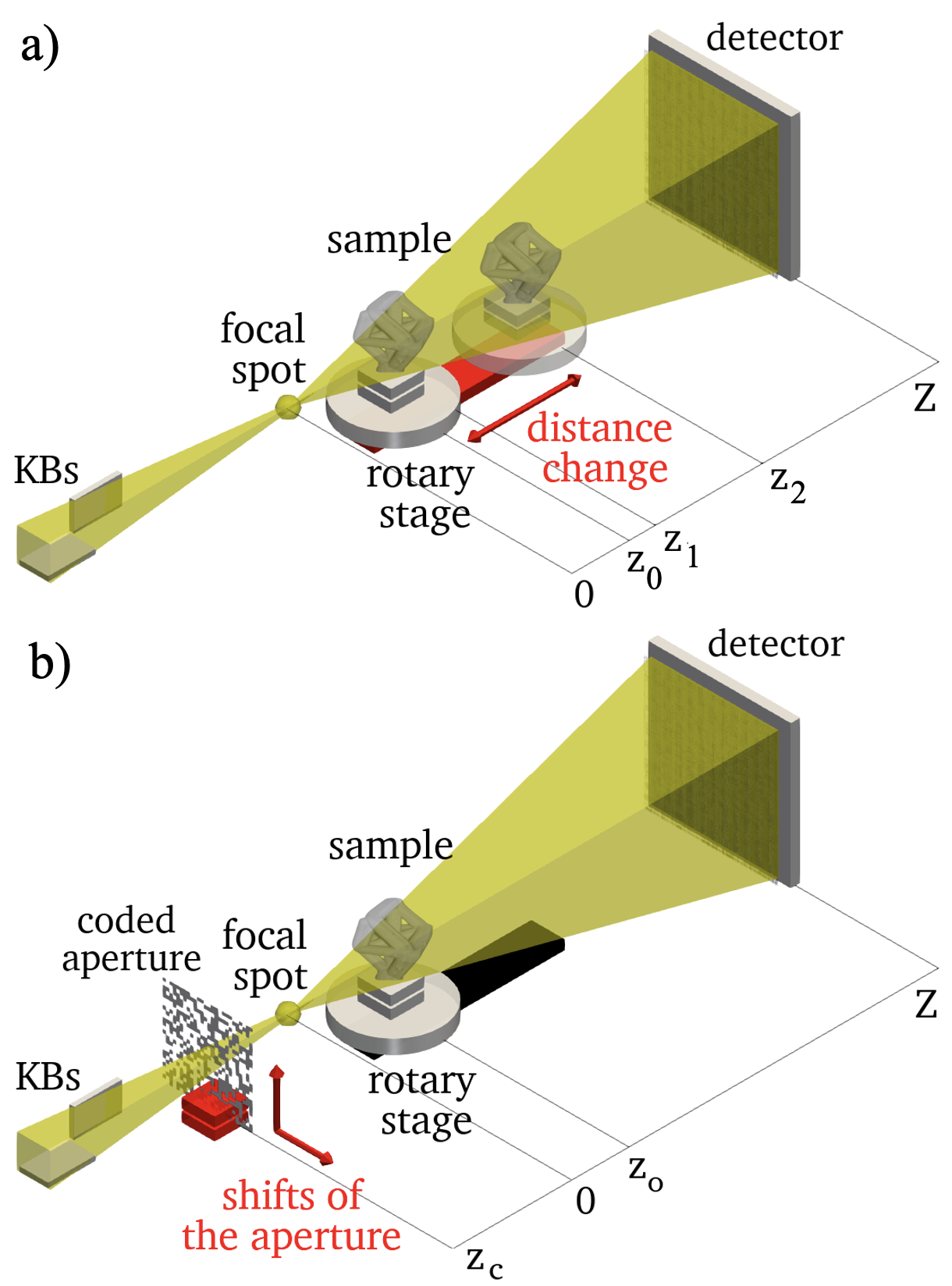}
\caption{Conventional nano-holotomography involves rotating the sample at multiple distances (a), while our proposed method rotates the sample at a single distance, shifting the coded aperture for each projection angle (b).}
\label{fig:scheme}
\end{figure}

Nano-holotomography is an advanced imaging technique that fuses X-ray tomography with phase-contrast imaging~\cite{cloetens1999holotomography}, allowing for high resolution visualization of internal structures, especially for materials with low absorption contrast. The technique is typically implemented with a Projection X-ray Microscope (PXM), where a coherent X-ray beam focused by Kirkpatrick–Baez (KB) mirrors produces magnified projections of a sample, see Fig.~\ref{fig:scheme}a. With the continuous improvement of the coherent synchrotron sources and the increasing efficiency of optical elements, nano-holotomography has become a leading tool for high spatial and temporal resolution~\cite{martinez2016id16b, kuan2020dense, robisch2020nanoscale, kalbfleisch2022x}. Moreover, PXM’s ability to operate at high X-ray energies (20-30keV)~\cite{martinez2016id16b,da2017high} allows for greater penetration power while minimizing radiation doses compared to other X-ray imaging techniques. 

Although new-generation synchrotrons provide the necessary flux and spatial coherence for high-resolution imaging, current PXMs worldwide face challenges in \textit{in-situ} imaging of dynamic systems, such as \textit{operando} battery studies and analysis of dynamic processes in biological/biomedical samples. Achieving high-resolution phase-contrast 3D imaging with a PXM currently requires collecting tomograms at multiple sample-to-detector distances, see Fig.~\ref{fig:scheme}a, which reduces temporal resolution and renders \textit{operando} studies impractical. Additionally, shifting the sample along the beam complicates the reconstruction procedures, necessitating alignment of data collected for different distances, and interpolation to correct for distance-dependent magnification on the detector. These complexities hinder real-time imaging, which is critical for steering holotomography experiments, whether selecting regions of interest for higher resolution scans or adjusting environmental conditions (pressure, temperature, charge) based on the real-time state of the sample.

Instead of acquiring tomographic data at multiple distances, we propose a novel approach: fixing the sample at a single distance and employing a continuously shifting coded aperture for each projection angle, positioned in front of the focal spot, as illustrated in Fig.~\ref{fig:scheme}b. This enables rapid imaging of samples and supports in situ or operando studies. The reconstruction scheme for the proposed coded nano-holotomography approach jointly solves the coded phase retrieval and 3D tomography problems using e.g.~the alternating direction method of multipliers (ADMM)~\cite{Boyd:11}. In what follows, we describe this new data acquisition scheme with coded apertures and demonstrate a corresponding reconstruction method for recovering 3D sample structures. Validation was performed with synthetically generated nano-holotomography data, utilizing the actual illumination (probe) function from the PXM instrument at the ID16A beamline of the ESRF. Our results are compared with the conventional multi-distance approach, demonstrating the efficiency and feasibility of our method for dynamic nano-holotomography experiments.


\textbf{Structured illumination with coded apertures:} Structured illumination is an effective technique for enhancing data collection efficiency, optimizing data quality, and minimizing radiation dose by utilizing patterned light. In this study, we demonstrate the use of binary coded apertures for nano-holotomography with structured illumination~\cite{forbes2021structured,gursoy2022depth}. These apertures, which are typically represented as binary matrix objects, can be fabricated using modern direct-write lithography and electroplating techniques. They are capable of producing features ranging from 1 to \SI{10}{\micro\metre} with an aspect ratio of up to 10.
Depending on the application, they can be made from high-Z materials such as gold or nickel, or from low X-ray absorbing materials such as SU-8 epoxy resin. We employ these apertures to selectively modulate the beam's phase, thereby enhancing the illumination structure rather than simply attenuating it. For instance, a \SI{5}{\micro\metre} thick gold coded aperture reduces the sample flux by only 5-10\% while significantly altering phase shifts at 20-\SI{30}{\keV} PXM energies. Using lower-Z materials would further minimize flux reduction. 


Interestingly, a version of the Fresnel scaling theorem applies even for objects placed before the focal spot, (as shown in \cite{nikitin2024x}, despite the fact that this widens the focal spot.) We have found that positioning the coded aperture upstream of the focal spot (as shown in Fig.~\ref{fig:scheme}b) is more effective than positioning it between the focal spot and the sample or downstream of the sample.

When the aperture is placed between the focal spot and sample, its patterns are magnified by the time they reach the sample plane, thereby not adding much new structure to the illumination pattern. Indeed, fabrication limitations restrict the minimum bit size, which may result in only a few bits being visible in the field of view and reduced interactions between the coded phase fringes and the sample. Conversely, placing the coded aperture downstream of the sample reduces the number of fringes due to the shorter beam propagation distance, thereby also not adding much new information on the detector.

To increase the independence of the data, we vary the illumination structure across different projection angles during tomography data acquisition. This is accomplished by synchronously shifting the coded aperture orthogonal to the beam while rotating the sample, as illustrated in Fig.~\ref{fig:scheme}b.

We have also found that the use of random patterns, such as sand paper, gives less new information than coded apertures. A reason for this could be that, after convolution with the Fresnel kernel, many smaller random features are simply blurred out and the final structured illumination in the object plane actually contains very little structure, thus giving little extra information between various positions of the aperture.

\textbf{Forward model for multi-distance nano-holotomography:} To lay the groundwork for our coded-aperture approach, we will begin by describing the baseline conventional multi-distance technique. The schematics of both techniques and beam propagation distances are given in Fig.~\ref{fig:scheme}. If we denote by $u=\delta+i\beta$ the complex refractive index of the object, then the transmittance function for this object can be calculated as $\psi_k=\exp(2\pi i \nu^{-1}\Rop_k u)$, where $\nu$ is the wavelength and $\Rop_k$ is the x-ray transform of the object for projection angle $\theta_k, \, k=0,\dots,N_{\theta}-1$. Also let $\Pop_z$ be the Fresnel propagator for distance $z$, and $\Mop_{m} f = f\left(x/m,y/m\right)$ be the magnification operator used for data scaling when modeling the conic beam propagation. 

For the conventional nano-holotomography model with propagation distances $z_{j}, \, j=0,\dots,N_z-1,$ and focus-detector distance $Z$ (see Fig.~\ref{fig:scheme}a) we refer to our derivations in~\cite{nikitin2024x}. In Appendix A of \cite{nikitin2024x} we provide an extension of the Fresnel scaling theorem, which transforms the conic beam propagation problem into an equivalent one in parallel beam geometry. The extended version involves double propagation for $j>0$: first between the probe $q$ given at the first sample plane $z_{0}$ and actual sample position $z_{j}$, and between the sample $z_{j}$ and detector $Z$. The model is given as follows
\begin{equation}\label{Eq:fwd_model_cone}
\begin{aligned}    
  &\left|\frac{z_0}{Z}\Mop_{m_0}\left(\Pop_{\zeta_j/\tilde m_j^2}\left(\Pop_{\omega_j}(q)\cdot\Mop_{1/\tilde m_j}(\psi_k) \right)\right)\right|^2 = d_{j,k}.
  \end{aligned}
\end{equation}
Here $\omega_j=(z_{j}-z_{0})/ m_j$, $\zeta_j=z_{j}(Z-z_{j})/Z$ are adjusted propagation distances,  $m_0=Z/z_0$ is the geometrical magnification on the detector from the first sample plane, and $ \tilde m_j=z_{j}/{z_{0}}$ is the magnification between position $z_0$ and $z_j$. Despite its complexity, the formula has a natural interpretation. Basically, it says that we first propagate the probe from the plane $z_0$ to the plane at $z_j$ by $\Pop_{\omega_j}(q)$, where it gets multiplied by a demagnified version of the object (since it occupies a smaller part of the beam when moved away from the focal spot). The result is then propagated to the detector by applying $(\lambda Z)^{-1}\Mop_{m_0}\Pop_{\zeta_j/\tilde m_j^2}$, which simply is an application of the Fresnel scaling theorem. 

\textbf{Forward model for coded nano-holotomography:} The proposed model with coded apertures can be built in a similar manner. Let $c$ be the transmittance function of the coded aperture. Assume the sample is placed at a distance $z_o>0$ and the coded aperture at a distance $z_c$ (which may be negative) with respect to the focal spot, as illustrated in Fig.~\ref{fig:scheme}b). The model still includes double propagation: between the coded aperture $z_c$ and sample $z_o$, and between the sample $z_o$ and detector $Z$. The proposed acquisition scheme assumes shifting the coded aperture (to change the beam structure) for each projection angle $k$, giving rise to a ``new'' cropped coded aperture for each angle, denoted by ${c}_k$. 
By switching to the parallel beam geometry following the formulas in Appendix A from \cite{nikitin2024x}, it can be verified that the forward model for coded nano-holotomography is
\begin{equation}\label{Eq:fwd_model_cone_code}
\begin{aligned}    
&\left|\frac{z_c}{Z}\Mop_{m_o}\Big(\Pop_{\zeta}\big(\Pop_{\omega}(\Mop_{m_c}(c_k\cdot {q}))\cdot \psi_k \big)\Big)\right|^2=d_k, 
  \end{aligned}
\end{equation}
where $m_o=Z/z_o$, $\zeta=(Z-z_o)/m_o$ and $\omega=(z_o-z_c) m_c$ are adjusted propagation distances, {${q}$ is the probe at the coded aperture position}, and $ m_c=z_{o}/{z_{c}}$ is the code magnification computed based on switching to new coordinates in parallel beam geometry. 

Again, despite the complicated appearance, the formula has a simple interpretation. First the code and the probe are propagated to the object-plane where they are properly magnified (by $\Mop_{m_c}$) before being multiplied with the object. The result is then propagated to the detector and magnified accordingly (by $\Mop_{m_o}$).
Note that the ``magnification'' $m_c$ becomes negative when $z_c<0$ and hence induces a flipping of the CA in the $x,y$ coordinates, and typically $|m_c|<1$ so in practice it is a demagnification of the coded aperture (making its patterns finer). 


\textbf{Solving the coded nano-holotomography problem:} The central part for reconstruction of nano-holotomography data collected with shifting coded aperture is solving the phase retrieval and tomography problems jointly. We implement the joint solver using the ADMM approach. We refer to ~\cite{Boyd:11,Li:15} for details about the ADMM, as well as to~\cite{Aslan:19, nikitin2019photon} where a similar approach was applied for solving the ptycho-tomography problem. Alternatively, one could also work directly with a gradient based approach, differentiating the cost functional with respect to $u$.

For simplicity, in this work we assume the probe function $q$ is known, although the algorithm can be extended by adding the probe reconstruction. The coded aperture function is assumed to be known, as it can be recovered before the experiment using techniques such as a scanning electron microscope or, preferably, near-field ptychography~\cite{stockmar2013near}. Near-field ptychography is particularly advantageous because its setup is compatible with that of holotomography, allowing for straightforward integration prior to data collection. Additionally, near-field ptychography provides a reliable estimate of both the illumination function and the coded aperture's transmittance function.

For the solution, we consider the minimization of 
\begin{equation}\label{eq:min1}
    \begin{aligned}    
  F(u)=\sum_{k=0}^{N_\theta-1}&\left\|\left|\Pop_{\zeta}\left(\Pop_{\omega}(\tilde c_k\cdot \tilde{q})\cdot e^{\frac{2\pi i}{\nu} \Rop_k u}) \right)\right|-\frac{z_c}{Z}\sqrt{\tilde d_k}\right\|_2^2
  \end{aligned}
\end{equation}
where $\tilde{c}_k$, $\tilde{q}$ and $\tilde{d_k}$ are suitably rescaled versions of the original functions. We remark that this rescaling is chosen so that, in practice, no interpolation is needed before multiplying objects belonging to the different planes.

\begin{figure}[ht!]
\centering\includegraphics[width=0.48\textwidth]{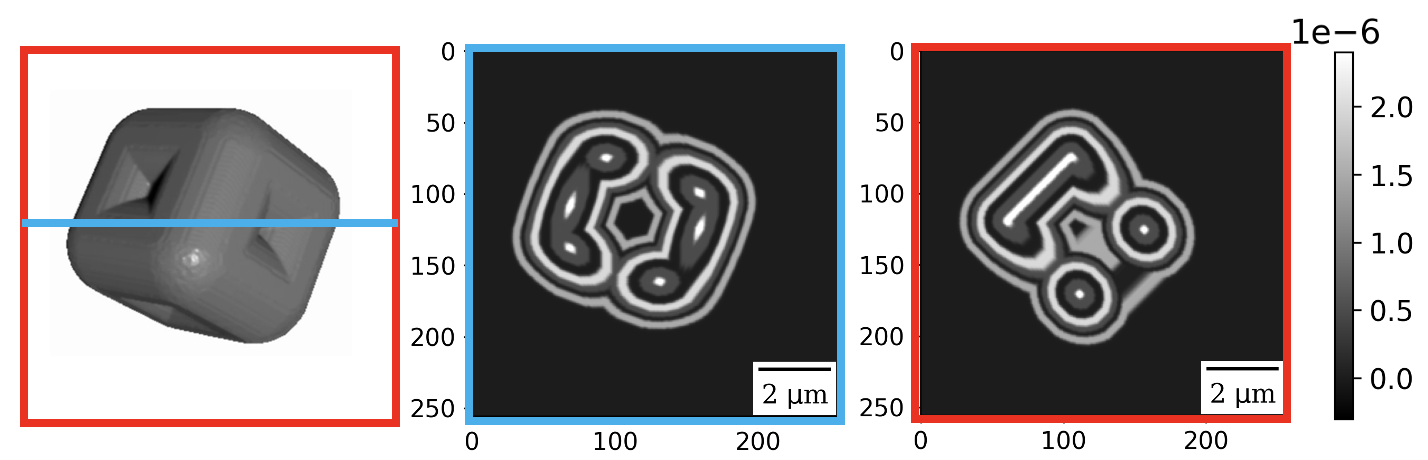}
\caption{Synthetic 3D sample used for simulations: 3D rendering and horizontal/vertical slices showing the layered structure. The colorbar corresponds to the real part ($\delta$) of the complex refractive index, the imaginary part ($\beta$) is 100 times lower.}
\label{fig:sample}
\end{figure}

By introducing auxiliary variables $\psi_k= e^{\frac{2\pi i}{\nu} \Rop_k u}$, the problem can be rewritten as a constraint minimization problem which can be solved with ADMM, which decomposes the problem into local subproblems related to $\psi_k$ and $u$. 
\textbf{Data Simulation:} For simulations we constructed a 3D sample which mimic a cubic structure containing thin layers of different refractive index, see Fig.~\ref{fig:sample}. The colorbar is associated with the phase component $\delta$, while the absorption component $\beta$ is 100 times lower. The whole volume size is $256\times256\times256$, where each voxel has size of \SI{40}{\nano\metre} in each dimension.

\begin{figure}[ht!]
\centering\includegraphics[width=0.45\textwidth]{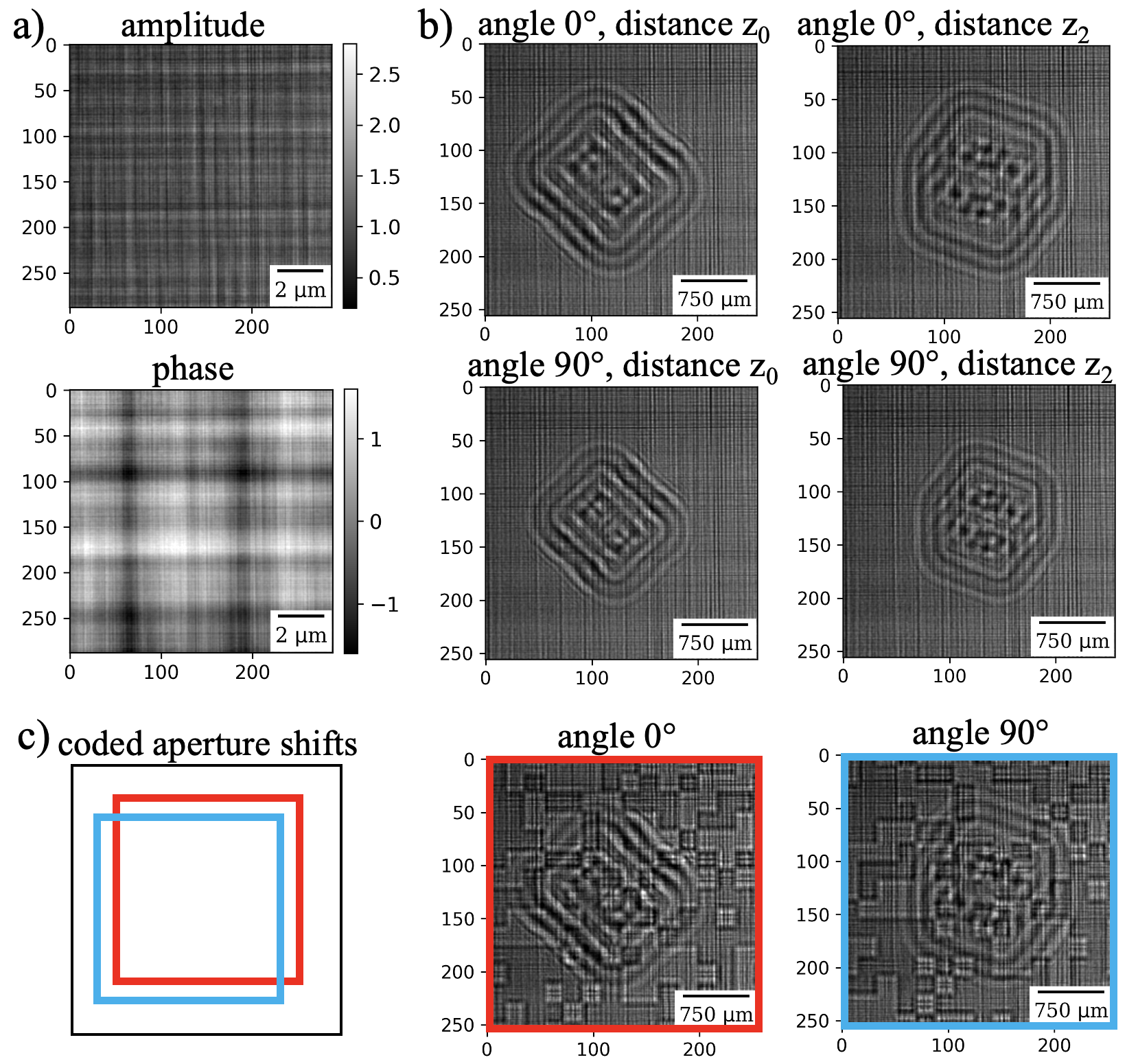}
\caption{Simulation of holotomography data with a real probe: the amplitude and phase of the probe at the first sample plane (a), examples of simulated data for the conventional multi-distance approach (b), examples of coded aperture shifts and corresponding simulated data (c). The colorbar for all images is identical to that of the probe amplitude image.}
\label{fig:data}
\end{figure}

For modeling holography data on the detector we partially employed the PXM instrument settings from beamline ID16A at the ESRF. The settings and parameters of the coded aperture are listed in Table~\ref{tab:pars}. To make the simulations more closely resemble real experimental conditions, we also utilized the real probe function $q$ of ID16A, recovered through the near-field ptychography. Fig.~\ref{fig:data}a shows the amplitude and phase of the probe. Note the structure of the probe contains the horizontal and vertical features from the multilayer coated KB mirrors. 

In our simulations, we use a practical data acquisition scenario with a gold coded aperture featuring a bit size of \SI{1}{\micro\metre} and a thickness of \SI{1.5}{\micro\metre}. The aperture is binary, with a semi-random pattern. The sample is positioned \SI{4.5}{\milli\metre} downstream of the focal spot, while the coded aperture is placed \SI{12}{\milli\metre} upstream of the focal spot.
In Fig.~\ref{fig:data}b we give examples of synthetically generated data using conventional nano-holotomography (\eqref{Eq:fwd_model_cone}), while Fig.~\ref{fig:data}c demonstrates images acquired with the coded aperture (\eqref{Eq:fwd_model_cone_code}).
\begin{table}[ht!]
    \centering
    \renewcommand{\arraystretch}{1.1}
    \small{
    \begin{tabular}{|c|c|c|c|}
    \hline
         Energy& \SI{33.35}{\keV} \\         
         Detector size after binning& $256\times256$ px\\
         Focal spot to detector distance & \SI{1.28}{\metre} \\          
         Focal spot to sample distances$^*$ & $4.214,\, 4.395,\, 5.118\SI{1}{\milli\meter}$ \\                  
         Focal spot to coded aperture &-\SI{12}{\milli\meter}\\
         Coded aperture parameters & material Au, bit size \SI{1}{\micro\meter}\\
         &thickness \SI{1.5}{\micro\meter}\\
         Voxel size for reconstruction & \SI{40}{\nano\metre}\\
         \hline
    \end{tabular}}\\
    $^*$ only the first distance for the coded aperture approach.
    \caption{Settings for nano-holotomography data simulations.}
    \label{tab:pars}
\end{table}

\begin{figure}[ht!]
\centering\includegraphics[width=0.48\textwidth]{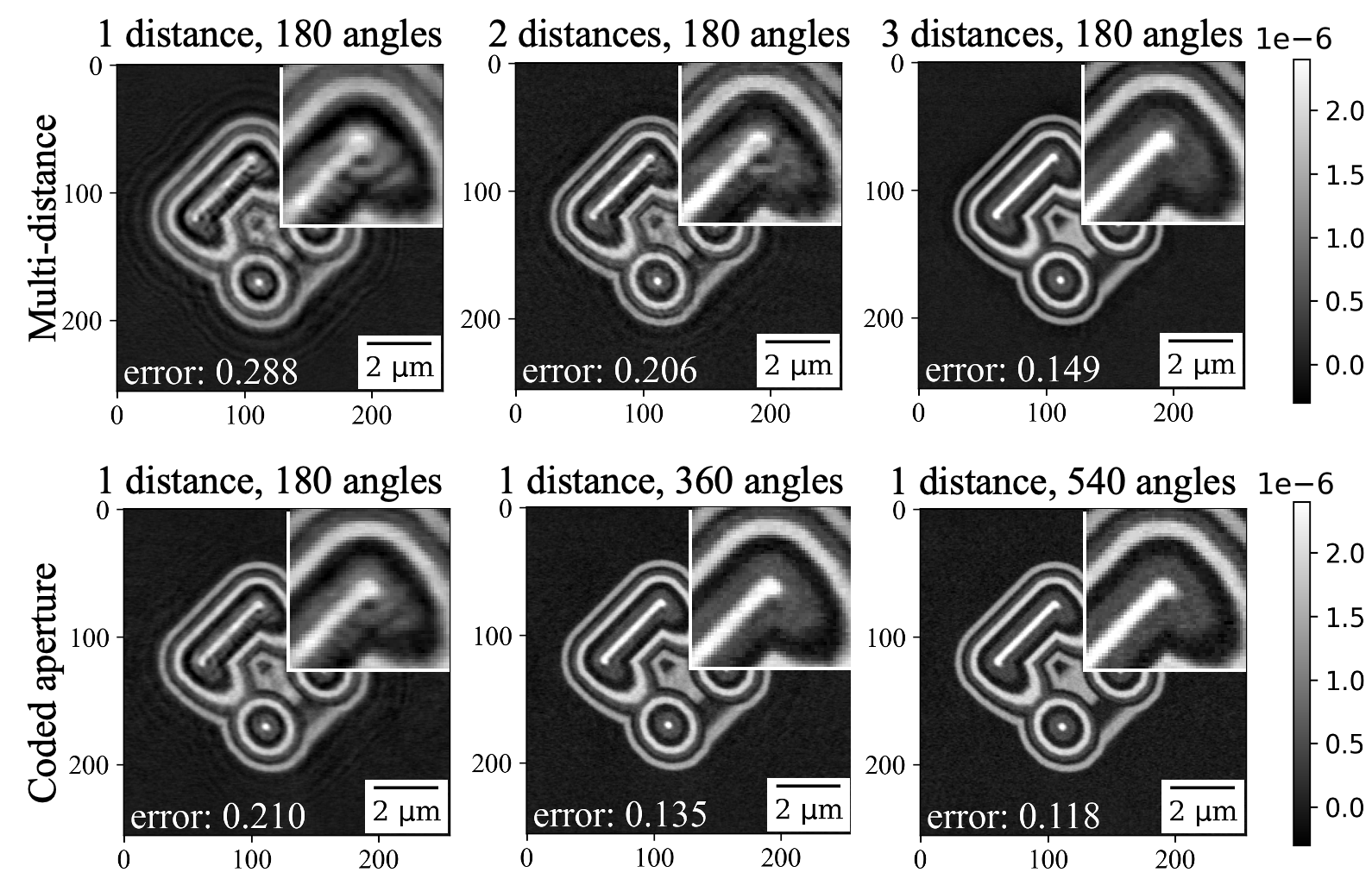}
\caption{Comparison between reconstructions of data simulated with the conventional multi-distance approach and with the approach utilizing single distance coded aperture approach.}
\label{fig:rec}
\end{figure}

\begin{figure}[ht!]
\centering\includegraphics[width=0.48\textwidth]{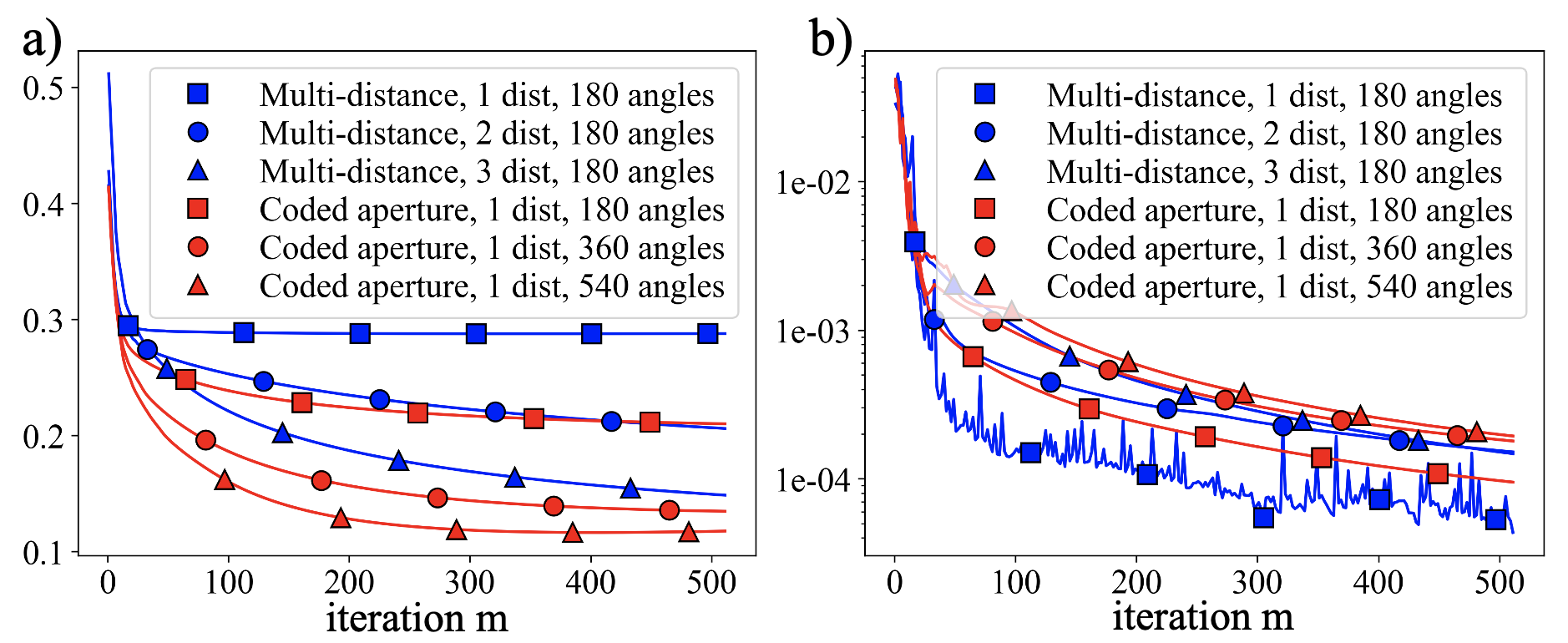}
\caption{Convergence analysis: a) convergence to the ground truth, $\|u^{(m)}-u^\text{GT}\|_2^2/\|u^\text{GT}\|_2^2$, and b) relative convergence $\|u^{(m+1)}-u^{(m)}\|_2^2/\|u^{(m)}\|_2^2$.}
\label{fig:rec_conv}
\end{figure}

\textbf{Comparison of reconstructions:} Fig.~\ref{fig:rec} presents the reconstruction results for both the conventional multi-distance approach and the proposed single-distance approach using coded apertures. Both methods employed a joint ADMM solver for holography and tomography. To assess robustness against noise, Poisson noise was assumed, simulating low radiation exposure per projection, which resulted in average photon counts ranging from 1000 to 2000 in the detector. While far-field propagation typically results in diffraction patterns with low photon counts, the near-field propagation characteristic of holotomography generally produces images that are affected by Poisson noise. 

In Fig.~\ref{fig:rec}, we show vertical slices through reconstructed real part $\delta$ of the object complex refractive index $u=\delta+i\beta$, with each slice accompanied by the normalized $L_2$-norm error compared to the ground truth. The first row shows results for the conventional holotomography method using 180 angles over a half-circle with 1, 2, and 3 acquisition distances. As expected, at least 3 distances are required for accurate reconstruction with this data collection approach. Results for 1 and 2 distances display poor quality, evident from the high $L_2$-norm errors. Increasing the number of projection angles does not improve the reconstruction quality, even with the joint ADMM solver.

In contrast, the second row of Fig.~\ref{fig:rec} demonstrates that the proposed coded holotomography method benefits from an increasing number of acquired projection angles. The shifted coded aperture introduces independent information for each angle, leading to improved reconstructions. Simulations with 180, 360, and 540 angles over a half-circle show that the proposed approach achieves lower errors with 540 angles compared to the 3-distance approach and delivers acceptable results with 360 angles. This reduction could potentially decrease the total radiation dose by 30\% in practical applications. 

Fig.~\ref{fig:rec_conv} further validates the efficiency of the proposed approach by demonstrating the convergence behaviour of the ADMM scheme for both approaches. The coded aperture approach exhibits a faster convergence compared to the conventional multi-distance approach. Notably, after 512 iterations, the coded aperture method with 180 or 360 angles reaches an error level comparable to the multi-distance approach with 2 or 3 distances, respectively, highlighting its potential for reduced radiation dose and scanning time in practical scenarios.

\textbf{Conclusions and Outlook:} We have developed and validated a new nano-holotomography method that employs coded apertures to structured beam illumination, with validation provided by numerical simulations. In contrast to conventional multi-distance approaches, which impose constraints on data collection time and introduce complexities related to sample alignment, our coded holotomography approach can recover a complex-valued object without requiring sample movement along the beam. This advancement offers enhanced stability and can facilitate efficient dynamic experiments in phase-contrast PXM, taking advantage of the coherence provided by new synchrotron sources globally. Additionally, we have demonstrated that the new approach can achieve high-quality reconstructions with fewer measurements, thereby has the potential to reduce the total radiation dose to samples. Looking forward, this technique promises significant advancements in dynamic and high-resolution imaging applications, especially in scenarios where reducing sample movement, conducting time-sensitive experiments, and radiation exposure is important. 


\section*{Acknowledgments}

This research used resources of the Advanced Photon Source, a U.S. Department of Energy (DOE) Office of Science user facility at Argonne National Laboratory and is based on research supported by the U.S. DOE Office of Science-Basic Energy Sciences, under Contract No. DE-AC02-06CH11357.

\section*{Disclosures}
The authors declare no conflicts of interest.

\bibliography{refs}

\end{document}